\documentclass[12pt]{article}
\usepackage[T1]{fontenc}
\usepackage[utf8]{inputenc}
\usepackage{lmodern, textcomp}
\usepackage{xcolor}
\usepackage{xspace}
\usepackage[english, francais]{babel}
\usepackage{amsmath, amsfonts, amssymb}
\usepackage[sort&compress,round]{natbib}
\usepackage{enumerate}
\usepackage{dsfont} \renewcommand{\mathbb}{\mathds}
\usepackage{graphicx, psfrag}
\usepackage[footnotesize, justification=centering]{caption}

\newcommand \Xset   {\mathbb{X}}
\newcommand \Rset   {\mathbb{R}}
\newcommand \Prob   {\mathsf{P}}

\newcommand \Esp    {\mathsf{E}}
\newcommand \II     {\mathcal{I}}
\newcommand \Xv     {\underline{X\mskip -1.5mu} \mskip 1mu}

\DeclareMathOperator \var  {var}
\DeclareMathOperator \argmin {argmin}

\newcommand \xtick[1] {\tiny \raisebox{-1mm}{#1}}

\begin{document}

\begin{center}
  {\large \sc 
    The Informational Approach to Global Optimization\\[-.5mm]
    in presence of very noisy evaluation results.\\[-.5mm]
    Application to the optimization of\\[0.5mm]
    renewable energy integration strategies
  }

\bigskip

\small
Héloïse Dutrieux$^{1,3}$ \& Ivana Aleksovska$^2$ \& Julien Bect$^{2}$ \\
\& Emmanuel Vazquez$^2$ \& Gauthier Delille$^1$ \& Bruno François$^3$

\bigskip

{\it %
  $^1$ Département \'Economie, Fonctionnement
  et \'Etudes des Systèmes Energétiques (EFESE), \\
  EDF R\&D, Clamart, France / \texttt{prenom.nom\at edf.fr}\\[0.5em]

  $^2$ Laboratoire des Signaux et Systèmes (L2S, UMR CNRS 8506) \\
  CentraleSupélec, CNRS, Université Paris-Sud, Gif-sur-Yvette, France \\
  \texttt{prenom.nom\at centralesupelec.fr}\\[0.5em]
 
  $^3$ Laboratoire d'\'Electrotechnique et d'\'Electronique de Puissance de
  Lille (L2EP) \\
  \'Ecole Centrale de Lille, Villeneuve d’Ascq, France
  / \texttt{prenom.nom\at ec-lille.fr}
}
\end{center}
\medbreak


{\bf R\'esum\'e.} Nous considérons le problème de l'optimisation globale d'une
fonction~$f$ à partir d'évaluations très bruitées. Nous adoptons un point de vue
bayésien séquentiel: les points d'évaluation sont choisis de manière à réduire
l'incertitude sur la position de l'optimum global de~$f$, cette incertitude
étant mesurée par l'entropie de la variable aléatoire correspondante
(Informational Approach to Global Optimization, Villemonteix et al.,
2009). Lorsque les évaluations sont très bruitées, l'erreur d'estimation de 
l'entropie par simulation conditionnelle devient non négligeable par rapport à
ses variations sur son domaine de définition. Nous proposons une solution à ce 
problème en choisissant les points d'évaluation comme si plusieurs évaluations 
allaient être faites en ces points. Une application à l'optimisation d'une stratégie 
d'insertion des énergies renouvelables dans un réseau de distribution d'électricité 
illustre la méthode proposée.

\smallskip

{\bf Mots-cl\'es.} Processus gaussiens;  Planification et analyse
d'expériences numériques; Optimisation bayésienne; \'Energies renouvelables;
Réseau de distribution électrique.

\selectlanguage{english}

\bigskip

{\bf Abstract.}  We consider the problem of global optimization of a
function~$f$ from very noisy evaluations. We adopt a Bayesian sequential
approach: evaluation points are chosen so as to reduce the uncertainty
about the position of the global optimum of $f$, as measured by the entropy of
the corresponding random variable (Informational Approach to Global
Optimization, Villemonteix et al., 2009). When evaluations are very noisy, the
error coming from the estimation of the entropy using conditional simulations
becomes non negligible compared to its variations on the input domain. We
propose a solution to this problem by choosing evaluation points as if several
evaluations were going to be made at these points. The method is applied to the
optimization of a strategy for the integration of renewable energies into an
electrical distribution network.

\smallskip



\selectlanguage{english}  \newpage

\section{Introduction}
\label{sec:intro}

Let~$f$ be a continuous real-valued function, defined on~$\Rset^d$ (or a subset
of~$\Rset^d$), $d \geq 1$. Given a finite set $\Xset \subset \Rset^d$, we consider
the problem of estimating the minimum $M = \min_{x\in\Xset} f(x)$ and the
corresponding set of minimizers, $x^{\star}\in \argmin_{x\in\Xset} f(x)$, using a
sequence of evaluations of $f$ at points $X_1, X_2, \ldots X_{n} \in \Xset$. In
this article, the evaluation results are assumed noisy: at each $X_i$, we
observe a perturbed value of $f(X_i)$. The construction of an optimization
algorithm $\Xv = \left( X_1, X_2, \ldots \right)$ is viewed as a sequential
decision problem: given $n$ (noisy) evaluation results at $X_1,\,\ldots,\,X_n$,
we must choose $X_{n+1}$ in order to get, in the end, the best estimators
of~$x^{\star}$ and~$M$ according to a certain loss function.

We adopt the following (classical) Bayesian approach for constructing
$\Xv$. The unknown function~$f$ is considered as a sample path of a
Gaussian random process $\xi$ defined on some probability space
$(\Omega, \mathcal{B}, \Prob_0)$, with parameter $x \in \Xset$. Then, a noisy
evaluation of $f$ at $X_i\in\Xset$ is modeled by the random variable
$\xi_i^{\rm obs}:=\xi(X_i)+\varepsilon_i$, $i=1,\,2,\,\ldots\,$, with
$\varepsilon_1,\,\varepsilon_2 \ldots \stackrel{\text{\tiny i.i.d}}{\sim}
\mathcal{N}(0, \sigma^{2})$ (here, $\sigma^2$ is assumed to be known). Denote
by $\Prob_n$ the conditional distribution $\Prob_0(\,\cdot \mid \II_n)$, where
$\II_n = \left\{ X_1,\, \xi_1^{\rm obs}\, \ldots,\, X_n,\, \xi_n^{\rm obs} \right\}$, 
and by $\Esp_n$ and $\var_{n}$
the conditional expectation $\Esp(\,\cdot\mid \II_n)$ and conditional
variance $\var(\,\cdot\mid \II_n)$ respectively.
Following~\citet{villemonteix:2009:IAGO} 
and~\citet{vazquez08:_global_optim_based_noisy_evaluat}, the efficiency of
an algorithm~$\Xv$ after~$n$ evaluations is measured using the posterior Shannon entropy
\begin{equation}
  \label{eq:entropy-loss}
  H(x^{\star};\II_n) = - \sum_{x\in\Xset} \Prob_n(x^{\star} = x)
  \log \Prob_n(x^{\star} = x)\,,  
\end{equation}
which quantifies the residual uncertainty about the position of
$x^{\star}$. Then, each new evaluation
point is chosen using a \emph{Stepwise Uncertainty Reduction} (SUR)
approach, which consists in minimizing a sampling criterion~$J_n$
that corresponds to the expected residual uncertainty on~$x^{\star}$ after $n+1$ evaluation results:
\begin{equation}
  \label{eq:jn} 
  X_{n+1} = \argmin_{x\in\Xset} J_n(x)
  \quad \text{with } \quad
  J_n(x):=\Esp_{n} \left( H(x^{\star};\II_{n+1}) \mid
    X_{n+1}=x \right). 
\end{equation}
Notice that $J_n(x)$ is an expectation with respect to the random
evaluation result $\xi_{n+1}^{\rm obs}$ at $X_{n+1}=x$. Minimizing $J_n$
is equivalent to maximizing the mutual information between~$x^{\star}$
and $\xi_{n+1}^{\rm obs}$. The reader is referred to \citet{picheny:2013:smo}
to a review of other sampling criteria for noisy optimization.

From a numerical point of view, the computation of $J_n$ is based on two
approximations. A first approximation is required for the computation of
the expectation in~(\ref{eq:jn}) with respect to the posterior distribution of
$\xi_{n+1}^{\rm obs}$ at $X_{n+1}=x$. Since $\xi$ and the evaluation
noise are Gaussian, the expectation in~(\ref{eq:jn}) is a one-dimensional integral with
respect to the Gaussian posterior density of $\xi_{n+1}^{\rm obs}$,
which can be carried out with a standard Gauss-Hermite quadrature. A
second approximation is needed to compute the entropy of the posterior
distribution of $x^{\star}$. \citet{villemonteix:2009:IAGO} estimate this
entropy by plugging into~(\ref{eq:entropy-loss}) an
estimator of $\Prob_n(x^{\star} = x)$, with $x$ ranging over $\Xset$, which,
in turn, is estimated by Monte-Carlo simulations of sample paths of
$\xi$ conditioned on $\II_n$.

When evaluations are noise-free, it is often possible to obtain a
satisfactory estimator of the entropy with a moderately large number of
sample paths ($\approx$ 1000). However, when the evaluation noise
becomes large, it appears that, for the same moderately large number of
sample paths, the variance of estimation of the entropy becomes non
negligible with respect to the information provided by a \emph{single}
evaluation. Then, minimizing $J_n$ to choose new evaluation points
becomes questionable. In this article, we propose to circumvent this
problem with a new sampling criterion where, in essence, we pretend that
several evaluations are going to be carried out instead of a single one.

\section{The Informational Approach to Global Optimization
  with (very) noisy evaluations}

\begin{figure}
  \begin{center}
    \def \wi {35.5mm}
    \psfrag{costMEuros}[cb][cb]{%
      \raisebox{4mm}{\scriptsize $J_n'(x)$}}
    \psfrag{tanPhiMin}[t][b]{%
      \raisebox{-3mm}{\scriptsize $x$}}
    \psfrag{x01}[ct][ct]{\xtick{$-1$}}
    \psfrag{x02}[ct][ct]{\xtick{$-0.9$}}
    \psfrag{x03}[ct][ct]{\xtick{$-0.8$}}
    \psfrag{x04}[ct][ct]{\xtick{$-0.7$}}
    \psfrag{x05}[ct][ct]{\xtick{$-0.6$}}
    \psfrag{x06}[ct][ct]{\xtick{$-0.5$}}
    \psfrag{x07}[ct][ct]{\xtick{$-0.4$}}
    \psfrag{x08}[ct][ct]{\xtick{$-0.3$}}
    \psfrag{x09}[ct][ct]{\xtick{$-0.2$}}
    \psfrag{x10}[ct][ct]{\xtick{$-0.1$}}
    \psfrag{x11}[ct][ct]{\xtick{$0$}}
    \psfrag{j01}[cr][cr]{\tiny $2.5$}
    \psfrag{j02}[cr][cr]{\tiny $2.6$}
    \psfrag{j03}[cr][cr]{\tiny $2.7$}
    \psfrag{j04}[cr][cr]{\tiny $2.8$}
    \psfrag{j05}[cr][cr]{\tiny $2.9$}
    \hspace{2mm}
    \includegraphics[width=38mm]{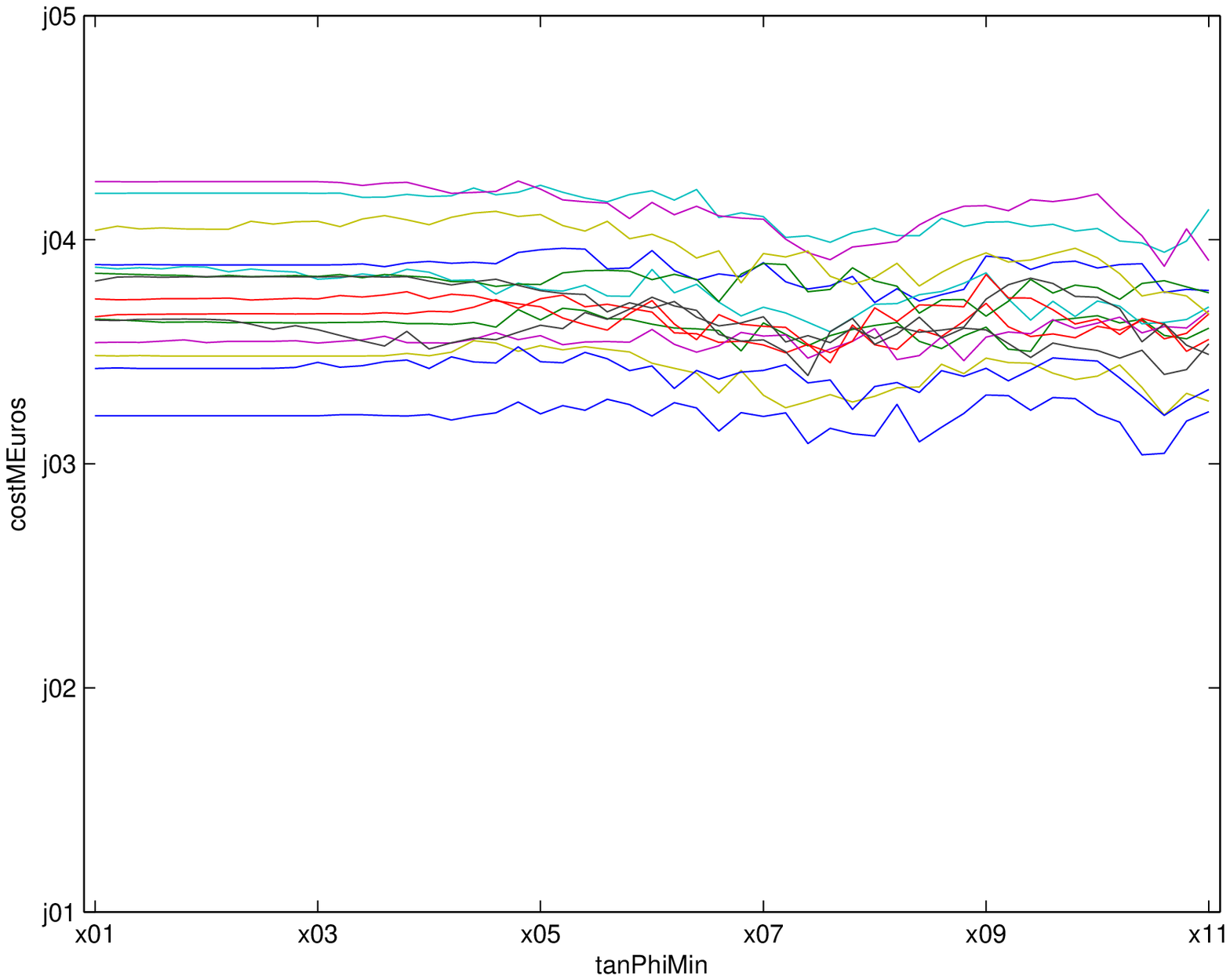}
    \hspace{1mm}
    \includegraphics[width=\wi]{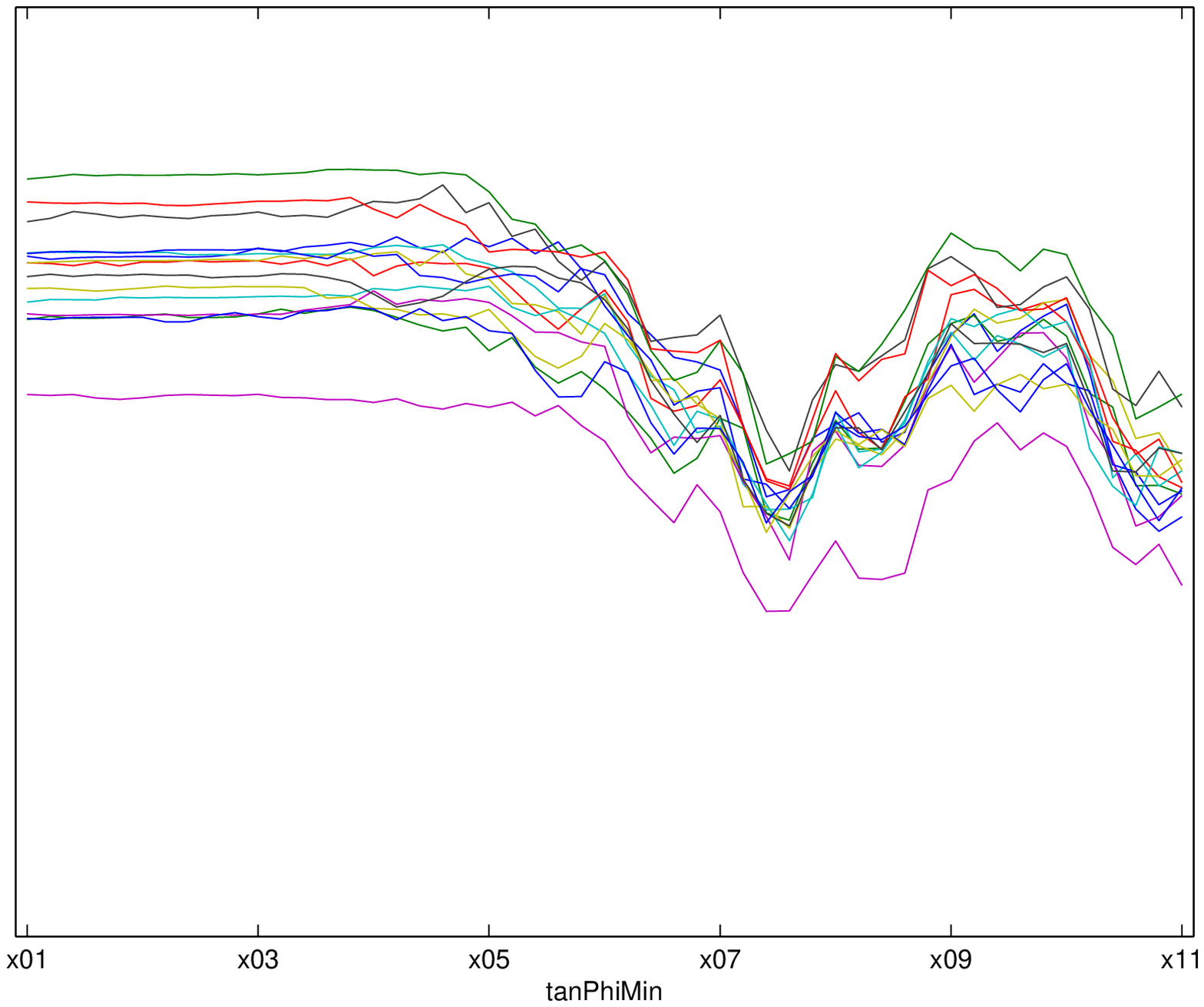}
    \hspace{1mm}
    \includegraphics[width=\wi]{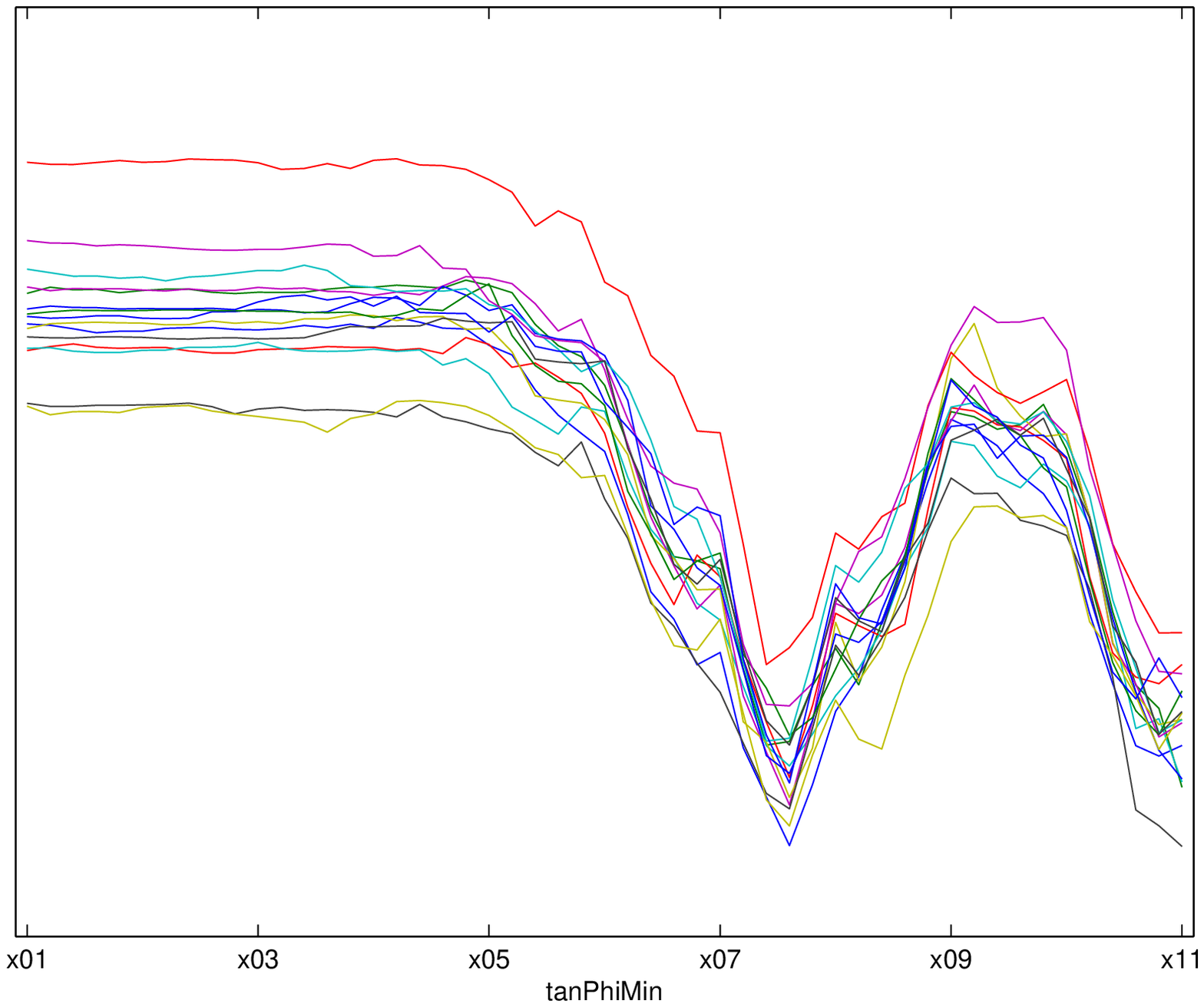}
    \hspace{1mm}
    \includegraphics[width=\wi]{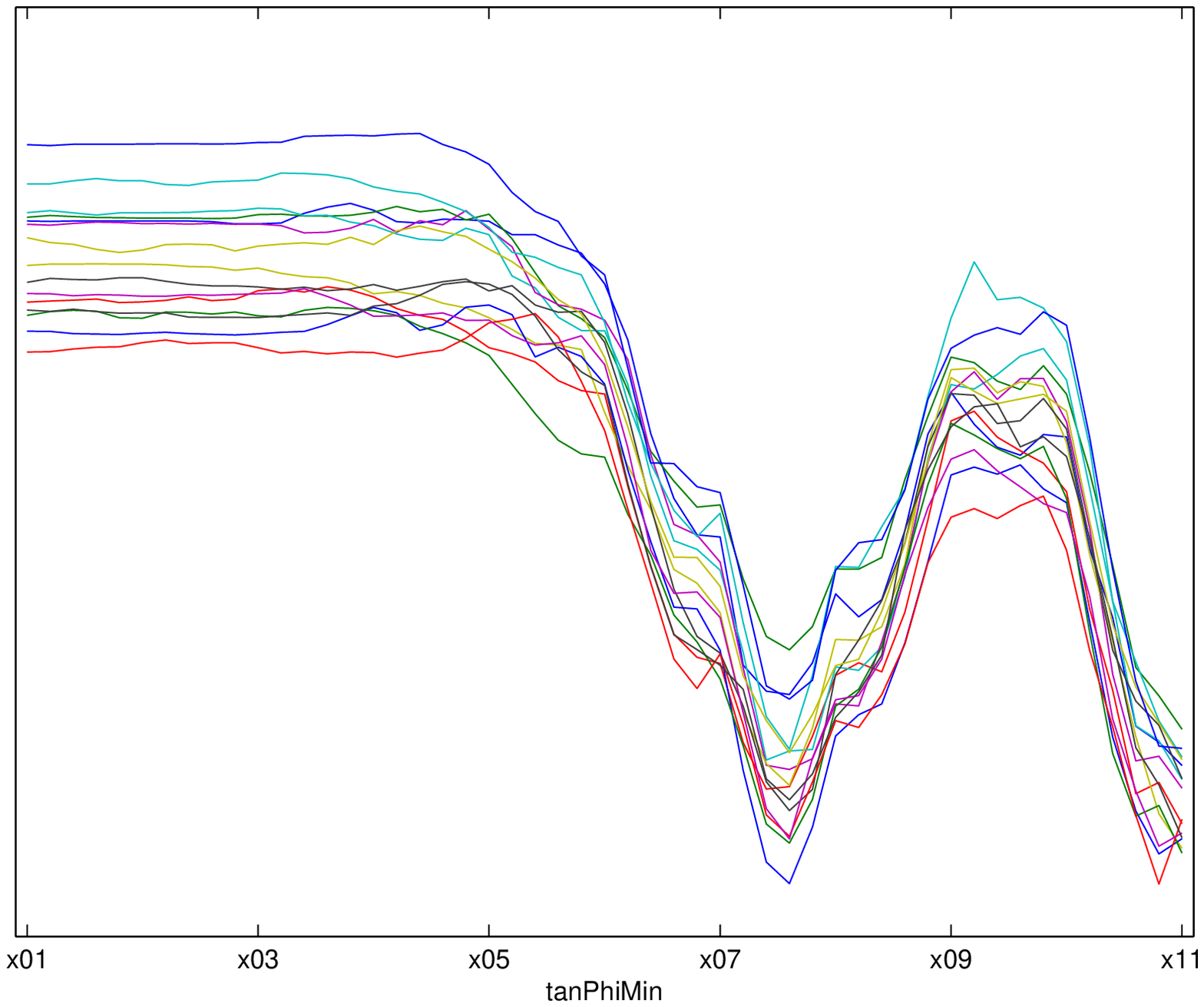}
  \end{center}
  \caption{Realizations of the numerical estimate of the sampling
    criterion~$J_n'(x)$ for the data shown in Figure~\ref{fig:appli-1}
    (right). Each figure represents $15$~independent realizations (corresponding
    to independent samples of conditional simulations). The batch size is, from
    left to right: $K = 1, 10, 100$ and~$+\infty$. A standard 15-order
    Gauss-Hermite is used for the integration and 1000 conditional samplepaths.
  }
  \label{fig:the-1}
\end{figure}

Since a single noisy evaluation provides limited information about $x^{\star}$,
and therefore yields by itself little progress in the optimization procedure,
the variations of~$J_n$ on~$\Xset$ can be dominated by its estimation error (as
illustrated in Figure~\ref{fig:the-1}, first left).

A natural idea to gain more information from noisy evaluations is to perform
several evaluations at each iteration of the optimization algorithm. Our
contribution is as follows: we suggest to build a sampling criterion
$J^{\prime}_n$ such that for all $x\in\Xset$, $J^{\prime}_n(x)$ corresponds to the
expected residual uncertainty about $x^{\star}$ when $K$ (noisy) evaluations of
$f$ are performed at $x$:
\begin{equation}
  \label{eq:jnp} 
  J^{\prime}_n(x):=\Esp_{n} \left( H(x^{\star};\II_{n+K}) \mid  X_{n+1}= \ldots = X_{n+K} =
    x \right). 
\end{equation}
The resulting criterion is illustrated in Figure~\ref{fig:the-1} with $K$ equal
to $10$, $100$ and~$+\infty$.

We refer to~$K$ as the \emph{virtual batch size}, since we do not actually
intend to perform~$K$ evaluations at the minimizer~$X_{n+1}$ of
$J^{\prime}_n$. Once~$X_{n+1}$ has been obtained by minimizing~\eqref{eq:jnp},
any number~$K_0$ of evaluations (between~$K_0 = 1$, as assumed in
Section~\ref{sec:intro}, and~$K_0 = +\infty$) can actually be performed at this
point; this number~$K_0$ is the \emph{actual batch size}. We suggest to take~$K$
large enough to make the error of estimation of $J^{\prime}_n$ small with
respect to the variations of the criterion, and to carry out only one actual
evaluation ($K_0 = 1$) at each iteration if evaluations are very expensive, or a
batch of size~$K_0 > 1$ (typically, $K_0 \ll K$) if evaluations are only
moderately expensive or if parallel processing is available. Another possibility
would be to update $K$ at each iteration so as to consider the whole remaining
budget of evaluations as suggested in~\citet{picheny:2010:iceo}.

The idea of considering $K$ evaluations \emph{at the same point}
in~(\ref{eq:jnp}) is only an artificial construction, motivated by the fact that
the numerical complexity of the computation of $J^{\prime}_n$ is the same as
that of~$J_n$. Indeed, it can be shown that the distribution of $\xi$
conditioned on $\II_{n+K}$ only depends in this case on $\xi_1^{\rm
  obs},\,\ldots,\,\xi_n^{\rm obs}$ and $\bar \xi_{n+1} = \frac{1}{K}\sum_{k=1}^K
\xi_{n+k}^{\rm obs} = \xi(x) + \frac{1}{K}\sum_k \varepsilon_{n+k}$. This has
two consequences. First, the expectation in~(\ref{eq:jnp}) is simply a
one-dimensional integral with respect to the (conditional) distribution of $\bar
\xi_{n+1}$, which is Gaussian, with mean equal to $\Esp_n(\xi(x))$ and variance
equal to $\var_{n}(\xi(x)) + \frac{1}{K}\sigma^2$. Second, the simulation of
sample paths of $\xi$ conditioned on the $n+K$ random variables $\xi_1^{\rm
  obs},\, \ldots,\, \xi_{n+K}^{\rm obs}$ boils down to the simulation of sample
paths of $\xi$ conditioned on the $n+1$ random variables $\xi_1^{\rm obs},\,
\ldots,\, \xi_{n}^{\rm obs}, \bar\xi_{n+1}$.

The optimization algorithm with the new criterion $J^{\prime}_n$ is available
for testing in a development branch of the STK toolbox \citep{stk}.

\section{Application}

The method is applied to the optimization of a strategy for the integration of
Renewable Energy Sources (RES) into an electrical distribution network. This
strategy describes how the Distribution System Operator (DSO) connects new
producers to the network under strict economic, safety and regulatory
requirements \citep{heloise:cired, heloise:powertech}. Our objective is to find
the optimal value of one parameter of the strategy, $x \in \left[ -1; 0
\right]$, so as to minimize the mean global cost of integrating about 20
megawatts of RES over 10 years.

\begin{figure}
  \begin{center}
    \psfrag{coutKEuros}[cb][cb]{%
      \raisebox{2mm}{\scriptsize cost}}
    \psfrag{coutMEuros}[cb][cb]{%
      \raisebox{2mm}{\scriptsize cost}}
    \psfrag{tanPhiMin}[t][b]{%
      \raisebox{-2mm}{\scriptsize $x$}}
    \psfrag{x01}[ct][ct]{\xtick{$-1$}}
    \psfrag{x02}[ct][ct]{\xtick{$-0.9$}}
    \psfrag{x03}[ct][ct]{\xtick{$-0.8$}}
    \psfrag{x04}[ct][ct]{\xtick{$-0.7$}}
    \psfrag{x05}[ct][ct]{\xtick{$-0.6$}}
    \psfrag{x06}[ct][ct]{\xtick{$-0.5$}}
    \psfrag{x07}[ct][ct]{\xtick{$-0.4$}}
    \psfrag{x08}[ct][ct]{\xtick{$-0.3$}}
    \psfrag{x09}[ct][ct]{\xtick{$-0.2$}}
    \psfrag{x10}[ct][ct]{\xtick{$-0.1$}}
    \psfrag{x11}[ct][ct]{\xtick{$0$}}
    \psfrag{y01}[cr][cr]{\tiny $0$}
    \psfrag{y02}[cr][cr]{\tiny $0.5$}
    \psfrag{y03}[cr][cr]{\tiny $1.0$}
    \psfrag{y04}[cr][cr]{\tiny $1.5$}
    \psfrag{y05}[cr][cr]{\tiny $2.0$}
    \psfrag{y06}[cr][cr]{\tiny $2.5$}
    \psfrag{y07}[cr][cr]{\tiny $3.0$}
    \includegraphics[width=7cm, height=45mm]{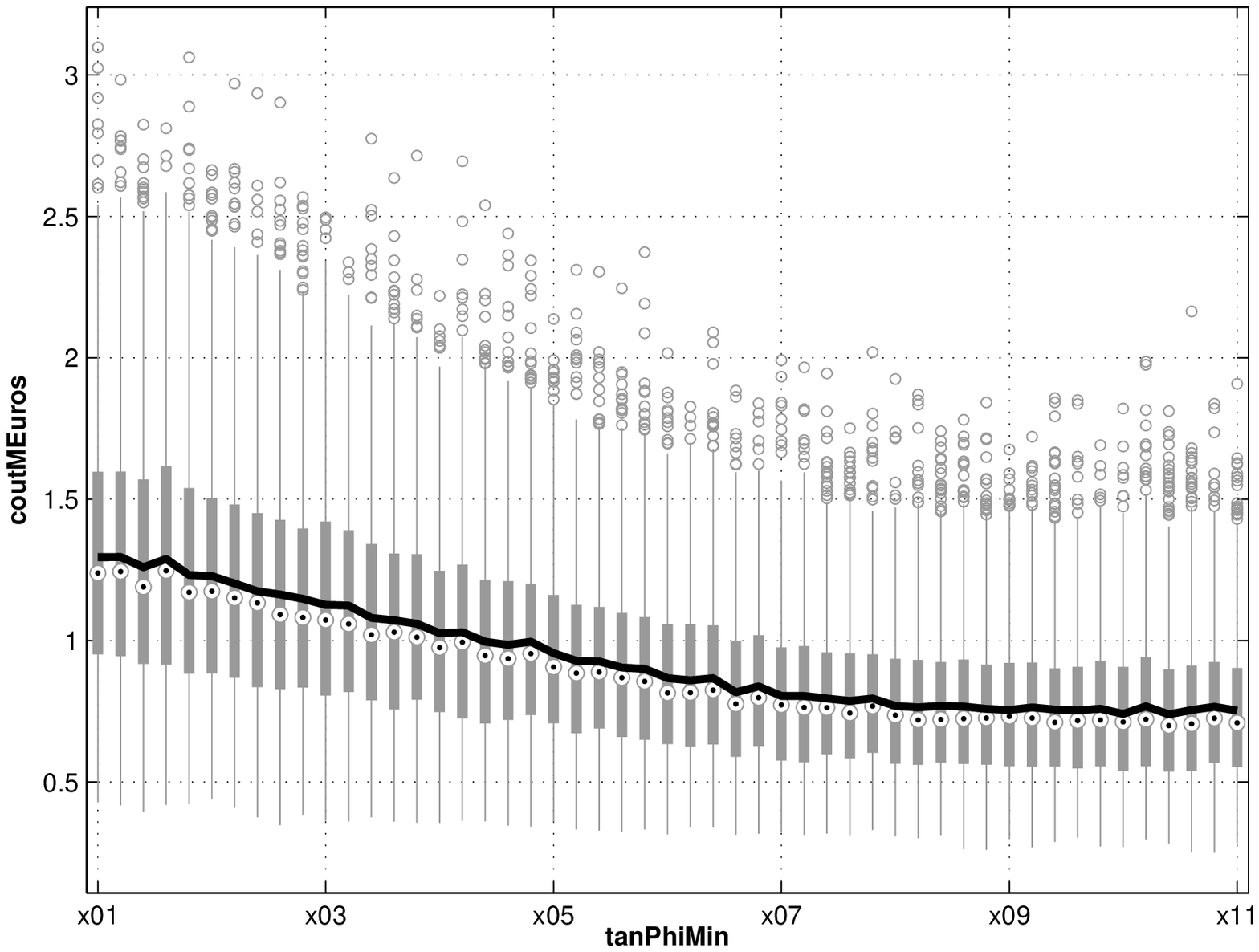}
    \hfill
    \raisebox{-1mm}{%
      \includegraphics[width=7cm, height=47mm]{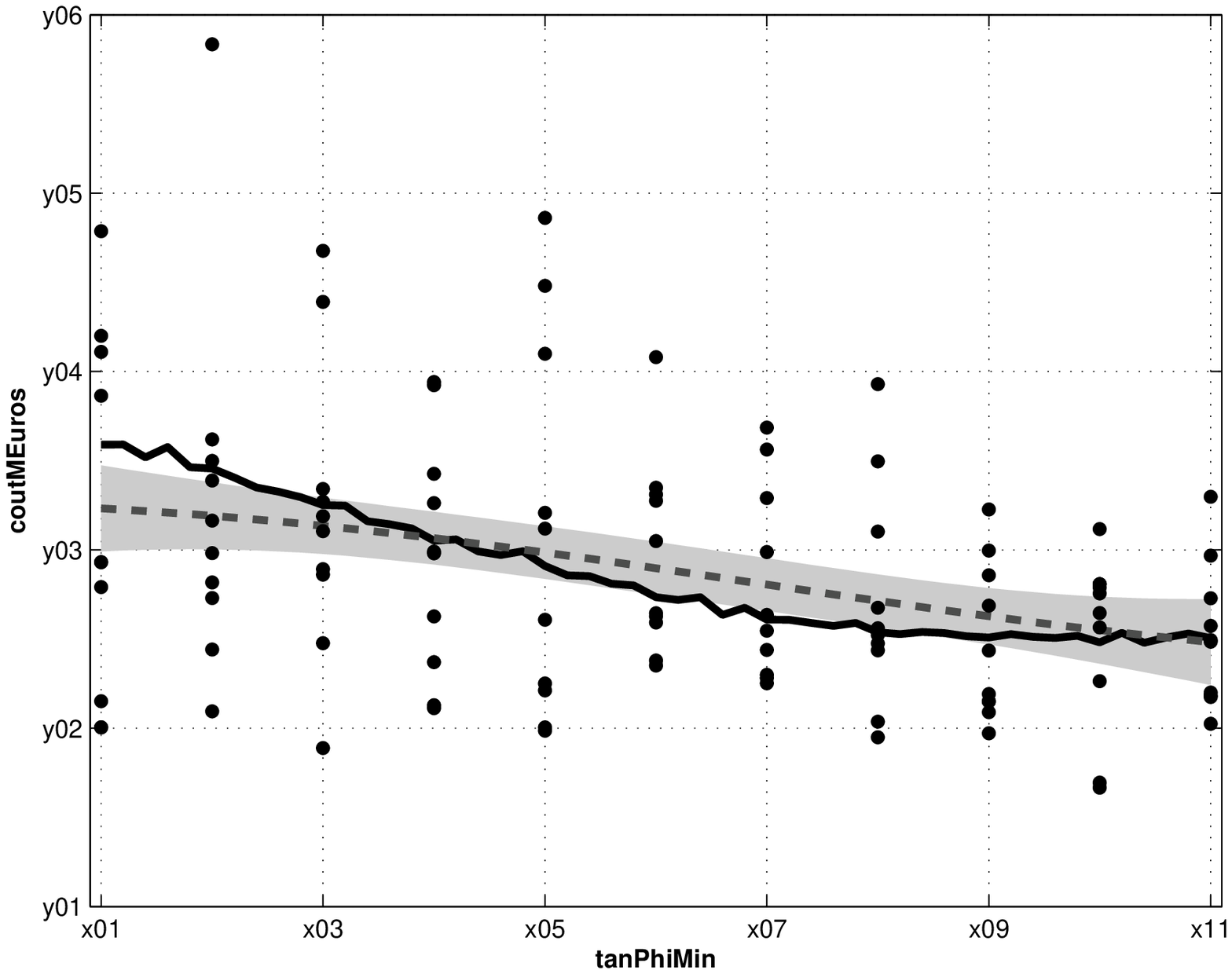}}
  \end{center}
  \caption{Left: Reference data. The search grid (on the $x$-axis) has $m = 51$
    points. On each point, approx.~1000 evaluation results are available. The
    solid black line represents the empirical mean. Right: initial sample of~$110$
    evaluations ($11$ batches of~$10$ evaluations). The dashed gray line represents the
    kriging mean. The grayed region represents pointwise credibility intervals
    with probability~$95\%$.}
  \label{fig:appli-1}
\end{figure}

The objective function is $f(x) = \Esp_S \left( C\left( x, S \right) \right)$,
where~$S$ denotes a 10-year scenario (consisting of several time series,
together with the characteristics of RES connection requests), $\Esp_S$ the
expectation with respect to a random scenario, and $C(x, S)$ the cost of the
strategy with parameter~$x$ applied to the scenario~$S$. The computation of~$C$
is performed by an expensive-to-evaluate computer program. We assume evaluations
of the form $\xi_{i}^{\rm obs} = C\left( X_i, S_i \right)$, where $S_1, S_2,
\ldots$ are independent scenarios generated by the same scenario generator (and
therefore identically distributed). This can be rewritten as $\xi_{i}^{\rm obs}
= f(X_i) + \varepsilon_i$, where the variables $\varepsilon_i = C\left( X_i, S_i
\right) - f \left( X_i \right)$ are independent and have zero mean. As shown in
Figure~\ref{fig:appli-1} (left), the evaluation results are very noisy in this
application. For the sake of simplicity, the noise variance is assumed to be
a known constant (estimated based on a few result evaluations) and
the variables $\varepsilon_i$, $i = 1, 2, \ldots$ will be assumed Gaussian.

We consider a budget of 2000 evaluations (without the initial sample) to find
the minimizer among 51 candidate points linearly spaced in $[-1, 0]$. A batch of
$K_0 = 10$ evaluations is performed at each iteration. We compare three ways of
constructing $\Xv$: using the sampling criterion $J^{\prime}_n$ when $K = K_0 =
10$ (denoted as IAGO $10$); using $J^{\prime}_n$ with $K = +\infty$ (denoted as
IAGO $+\infty$); and, as reference, choosing~$X_{n+1}$ at random, uniformly in
the set of candidate points (denoted as IID). The kriging model parameters are
firstly estimated on an initial sample of 110 evaluations (11 batches of 10
evaluations as shown in Figure~\ref{fig:appli-1}, right), then adjusted after
each new batch of evaluations.

\begin{figure}
  \newcommand \xOptEstim {\widehat{x^*_n}}
  \newcommand \MOptEstim {\widehat{M}_n}
  \begin{center}
    \def \wi {48mm}
    \psfrag{legend1111}{\tiny IID}
    \psfrag{legend2222}{\tiny IAGO $10$}
    \psfrag{legend3333}{\tiny IAGO $+\infty$}
    \psfrag{xopt}{\raisebox{3mm}{\footnotesize $\xOptEstim$}}
    \psfrag{yopt}{\raisebox{2mm}{\footnotesize $\MOptEstim$}}
    \psfrag{H}{\footnotesize $H_n$}
    \psfrag{xxx}[t][b]{{\footnotesize $n$}}
    \psfrag{x01}[ct][ct]{\xtick{$0$}}
    \psfrag{x02}[ct][ct]{\xtick{$500$}}
    \psfrag{x03}[ct][ct]{\xtick{$1000$}}
    \psfrag{x04}[ct][ct]{\xtick{$1500$}}
    \psfrag{x05}[ct][ct]{\xtick{$2000$}}
    \psfrag{y01}[cr][cr]{\tiny $-0.4$}
    \psfrag{y02}[cr][cr]{\tiny $-0.3$}
    \psfrag{y03}[cr][cr]{\tiny $-0.2$}
    \psfrag{y04}[cr][cr]{\tiny $-0.1$}
    \psfrag{y05}[cr][cr]{\tiny $0$}
    \includegraphics[width=\wi]{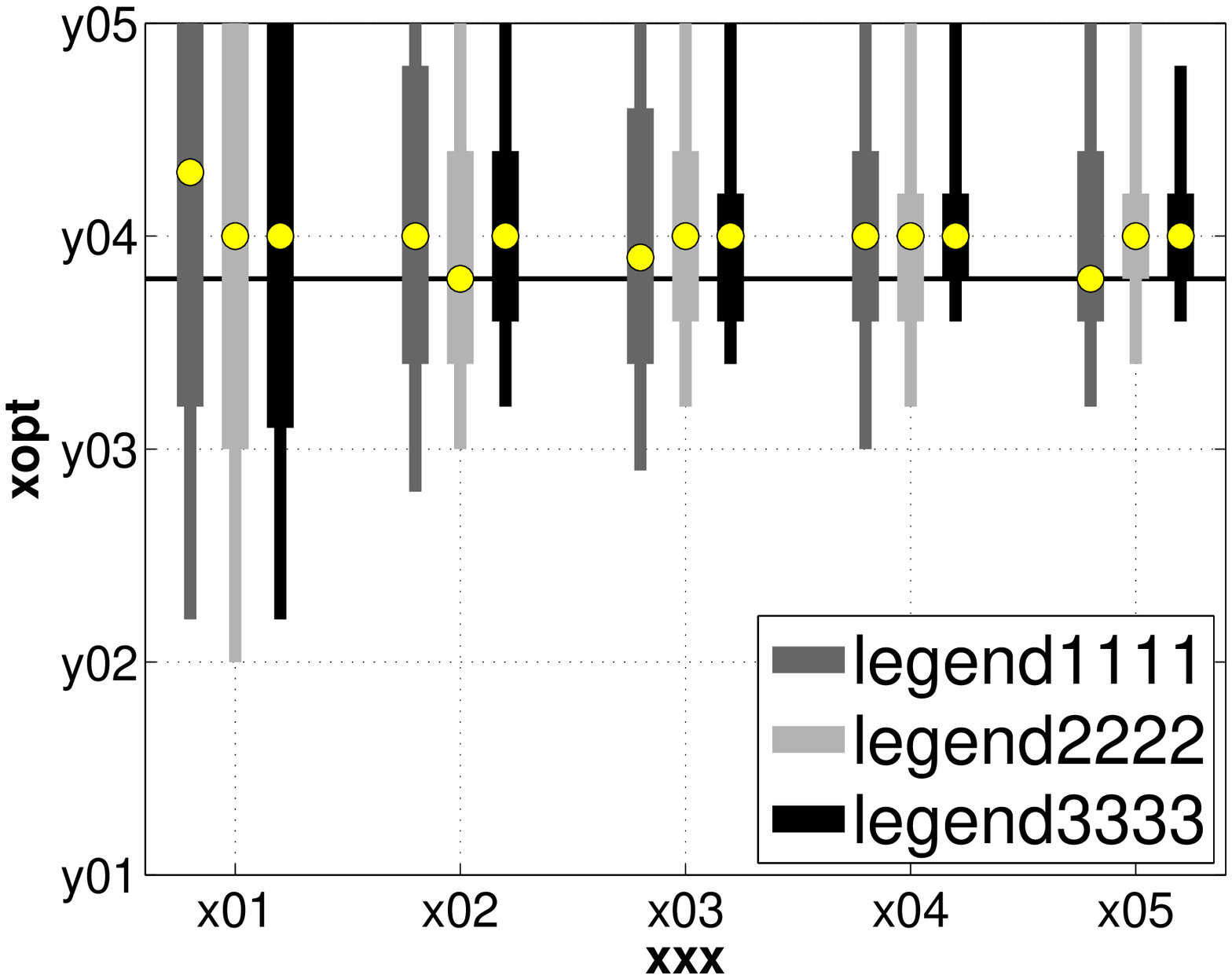}
    \hfill
    \psfrag{y01}[cr][cr]{\tiny $0.6$}
    \psfrag{y02}[cr][cr]{\tiny $0.65$}
    \psfrag{y03}[cr][cr]{\tiny $0.7$}
    \psfrag{y04}[cr][cr]{\tiny $0.75$}
    \psfrag{y05}[cr][cr]{\tiny $0.8$}
    \includegraphics[width=\wi]{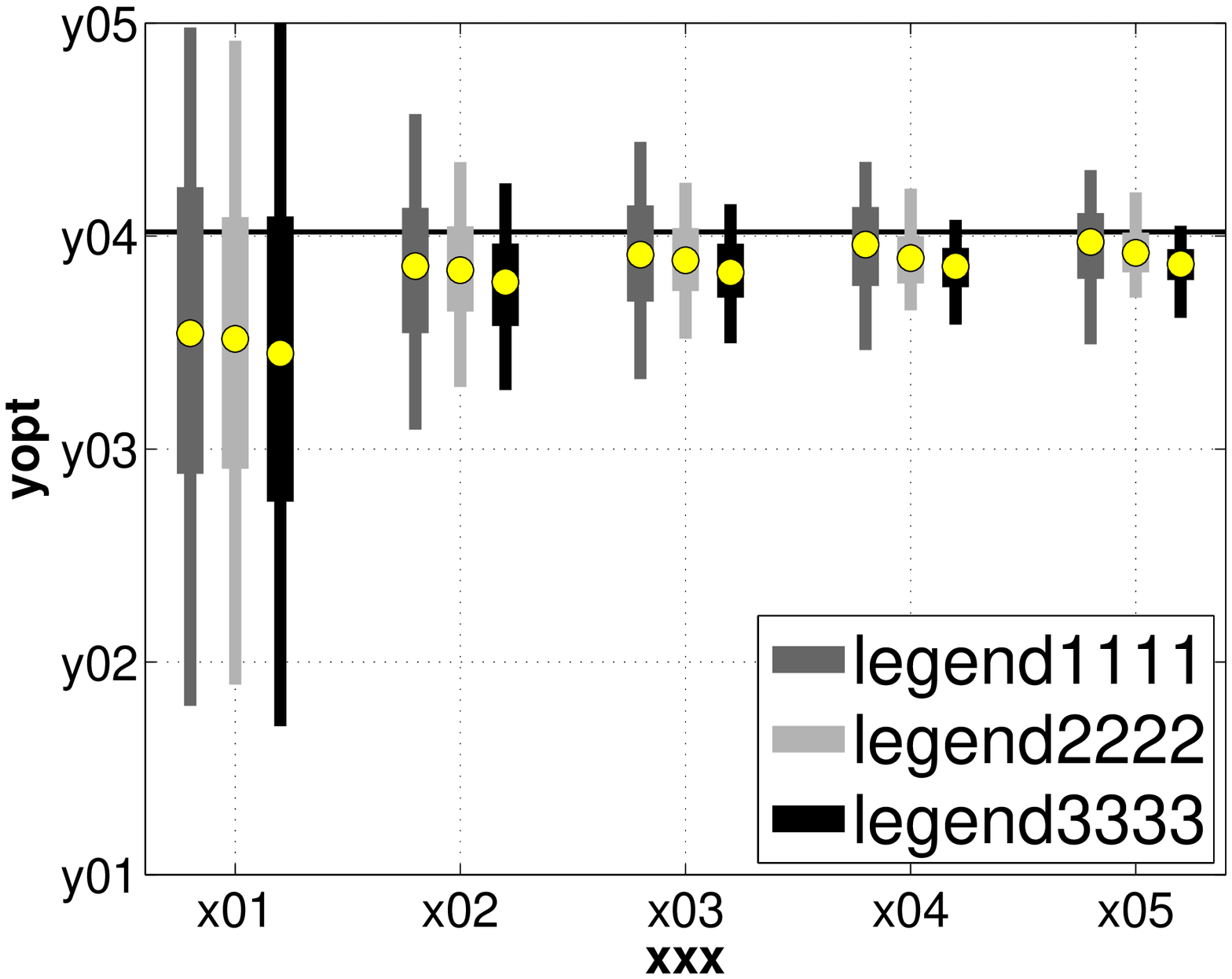}
    \hfill
    \psfrag{y01}[cr][cr]{\tiny $0$}
    \psfrag{y02}[cr][cr]{\tiny $1$}
    \psfrag{y03}[cr][cr]{\tiny $2$}
    \psfrag{y04}[cr][cr]{\tiny $3$}
    \includegraphics[width=\wi]{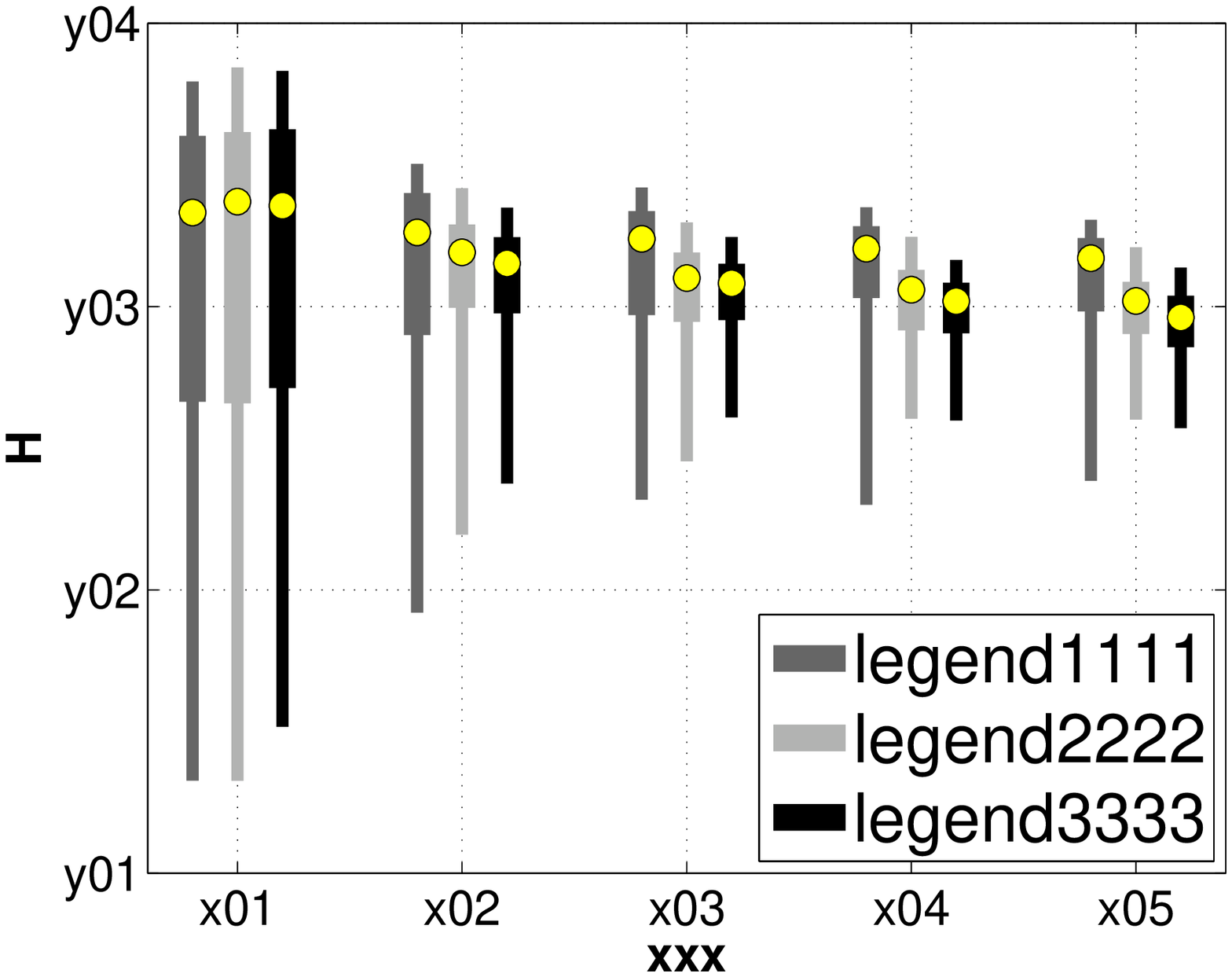}
  \end{center}  
  \caption{Distribution of the estimated minimizer~$\xOptEstim$\ (left), the
    estimated minimum~$\MOptEstim$ (center) and the posterior entropy~$H_n$ of
    the minimizer (right) over 500 optimization runs. On each box, the central
    mark is the median, the edges of the box are the 25th and 75th percentiles,
    and the whiskers are the 5th and 95th percentiles. The thick black lines
    indicate the value obtained on our reference dataset.}
  \label{fig:appli-2}
\end{figure}

Figure~\ref{fig:appli-2} depicts the distribution of the estimated minimizer,
the estimated minimum and the posterior entropy of the minimizer over the 500
optimization runs. IAGO $+\infty$ converges towards the area of interest faster
than IID and IAGO $10$. It is worth noting that a budget of 2000 evaluations
does not suffice to locate the minimizer accurately. In fact, even 1000
evaluations at each candidate point (as in Figure~\ref{fig:appli-1}, left),
would not locate it much more precisely (result not shown).

\section{Conclusion}

We have proposed a new sampling criterion for the problem of global optimization in
presence of very noisy evaluations, assuming that several evaluations are going to be made
at a new evaluation point (even if they are not in practice). The proposed method has
been applied to the optimization of a renewable energy integration strategy and shown
to outperform plain IID sampling and the original IAGO criterion.

\section{Acknowledgements}

This work was supported in part by the French National Research Agency (ANR) 
under PROGELEC 2012 grant program (APOTEOSE project, ref. ANR-12-PRGE-0012). 
The authors gratefully acknowledge its support.

\footnotesize \setlength{\bibsep}{0.08cm plus 0.05cm}
\bibliographystyle{plainnat}
\bibliography{jds15-iago}

\end{document}